\documentclass[a4paper,11pt]{article}
\pdfoutput=1 

\usepackage{jcappub} 

\usepackage[T1]{fontenc} 
\usepackage{amsmath}
\usepackage{scalerel}
\usepackage{tensor}
\usepackage{color, colortbl}
\definecolor{light-gray}{gray}{0.95}
\definecolor{darkred}{rgb}{0.7,0,0}

\usepackage[export]{adjustbox}
\usepackage{url}

\title{\Large \boldmath Analytic Correlation of Inflationary Potential to Power Spectrum Shape:\\
{\large Limits of Validity, and `No-Go' for Small Field Model Analytics}}

\author[]{Ira Wolfson}

\affiliation[]{International School for Advanced Studies (SISSA), Data Science Excellence Department, Via Bonomea 265, 34136 Trieste, Italy
}

\emailAdd{iwolfson@sissa.it}

\abstract{The primordial power spectrum informs the possible inflationary histories of our universe. Given a power spectrum, the ensuing cosmic microwave background is calculated and compared to the observed one. Thus, one focus of modern cosmology is building well-motivated inflationary models that predict the primordial power spectrum observables. The common practice uses analytic terms for the scalar spectral index $n_s$ and the index running $\alpha$, forgoing the effort required to evaluate the model numerically. However, the validity of these terms has never been rigorously probed and relies on perturbative methods, which may lose their efficacy for large perturbations. The requirement for more accurate theoretical predictions becomes crucial with the advent of highly sensitive measuring instruments.  This paper probes the limits of the perturbative treatment that connects inflationary potential parameters to primordial power spectrum observables. We show that the validity of analytic approximations of the scalar index roughly respects the large-field/small-field dichotomy. We supply an easily calculated measure for relative perturbation amplitude and show that, for large field models, the validity of analytical terms extends to $\sim 3\%$ perturbation relative to a power-law inflation model.
Conversely, the analytical treatment loses its validity for small field models with as little as $0.1\%$ perturbation relative to the small-field test-case. By employing the most general artificial neural networks and multinomial functions up to the twentieth degree and demonstrating their shortcomings, \textit{we show that no reasonable and universal analytic expressions correlating small field models to the observables they yield exists}.
Finally, we discuss the possible implications of this work and supply the validity heuristic for large and small field models.
}

\begin{document}
\maketitle
\flushbottom
\section{Introduction}
Ever since the late ’70s, the scenario of inflation \cite{Starobinsky:1980te,Sato:1980yn,Guth:1980zm,Linde:1981mu,Albrecht:1982wi} has increasingly become the best current explanation for our observations of the universe \cite{Bennett:1996ce,Hinshaw:2012aka,Akrami:2018odb}. As an epoch of accelerated expansion of the universe, inflation successfully explains the apparent isotropy and homogeneity of the Cosmic Microwave Background Radiation (CMB), even between space-like separated regions. Inflation predicts an almost scale-invariant curvature perturbation Primordial Power Spectrum (PPS) \cite{Mukhanov:1981xt,Hawking:1982cz,Starobinsky:1982ee,Guth:1982ec,Bardeen:1983qw} and negligible Non-Gaussianities \cite{Bartolo:2004if}. Perhaps the most sought-after data is a stochastic primordial Gravitational Waves (GW) signal \cite{Grishchuk:1974ny,Starobinsky:1979ty,Abbott:1984fp}. A detection of primordial GW is commonly regarded as proof-positive of the inflationary paradigm \cite{Lyth:1996im}. The absence of a GW signal detection at the current instrumental sensitivity levels implies an upper bound on the amplitude of such a signal \cite{Akrami:2018odb,Ade:2018gkx}. Additionally the Lyth bound \cite{Lyth:1996im} is thought to tightly constrict the slope of the inflationary potential and the inflaton field excursion during inflation, given a detection of a primordial GW signal. Consequently, until recently, it was thought that inflationary models in which the field excursion is of $\mathcal{O}(1)$ are observationally disfavored. The original taxonomy of models \cite{Dodelson:1997hr} differentiates between two types of models: 1) models in which the inflaton field is displaced far from some local minima  and rolls down the potential to the end of inflation at that minima; and 2) models in which the field's starting position is close to a maxima. The former are called `Large Field Models' (LFMs) whereas the latter are `Small Field Models' (SFMs). This taxonomy also reflects our preconceptions of the inflationary potential. Usually in LFMs we assume that the high energy behaviour is well understood and the entire model can be tuned by very few degrees of freedom. However, the SFM point of view states that the high energy behaviour is not specifically known, and we default to some Taylor expansion description of the potential. Usually the field excursion in LFMs is of the order of $\Delta\phi\sim\mathcal{O}(10)$ whereas in SFMs the field excursion is of the order of $\Delta\phi\lesssim\mathcal{O}(1)$. Thus the simplified taxonomy became a division between models with $\Delta\phi \gtrsim 10$, loosely called LFMs and models for which $\Delta\phi\lesssim\mathcal{O}(1)$, which are loosely referred to as SFMs. More details regarding different taxonomies can be found in \cite{Wolfson:2021utn}. Theoretically, recent developments in string theories (c.f. \cite{Palti:2019pca,Matsui:2018bsy,Grimm:2018ohb,Ben-Dayan:2018mhe}) and the lower energy regime Effective Field Theories (EFT) they produce, seem to strongly favor SFMs. \\

In building a viable inflationary model, the final test is whether or not the model yields the correct statistical properties we observe in the CMB. This process entails tracking the background geometry and other fields through inflation, calculating the curvature perturbations, decomposing these perturbations to different modes, and finally, deriving the PPS. Several of these stages can be complex or very time-consuming. Thus, an analytical tool was derived by way of perturbation theory over the inflationary potential itself \cite{Stewart:1993bc} to connect the functional shape of the model to the shape of the PPS.\\

Meanwhile, our observational instruments have gotten progressively better. We are now at the point where most of the cosmological observables are measured to a precision of better than $0.1\sim1\%$ relative error \cite{Planck:2018vyg}. Specifically the scalar index $n_s$ is measured to the a $\mathcal{O}(10^{-4})$ precision, which is $\sim 0.01\%$ relative error, and the index running $\alpha$ is measured to the same numerical precision. If we are to use CMB observations to gain insight into high energy physics, it is thus crucial that we use tools that yield results of the same precision. \\

In this work we probe the viability limits of this perturbative treatment, which was presented in \cite{Stewart:1993bc,Lyth:1998xn} and spawned numerous successors \cite{Lyth:1998xn,Dodelson:2001sh,Gong:2001he,Stewart:2001cd,Schwarz:2001vv}. While this perturbative treatment is well motivated and widely used, its limits of validity have never been rigorously probed. We show that when applied to LFMs, the perturbative method approximates the numerical results to a high degree of accuracy. However, for SFMs, the analytical approximation often fails spectacularly \cite{Wolfson:2016vyx,Wolfson:2018lel}. Thus, we already know that the analytical terms are incorrect in some capacity. With careless generalization, this may undo about thirty years worth of model-building and particle physics insights gleaned from inflationary cosmology. So, we aim to constrain this break-down of analytical expression by supplying limits of validity, expressed by a simple heuristic, yet quantitative test. The rationale is the following: If a model passes the test, it warrants no further proof for the validity of its analytical results. However, failing the test does not point to invalid results. Rather, it implies the model should undergo a more rigorous numerical test.\\

While the perturbative-analytical approach is widely applied, it is by no means the only method available. Most notably in \cite{Dvorkin:2009ne} the authors employ a numeric-perturbative approach where the eigenfunction problem of slow-roll inflation is solved. The Green's function used in that approach is the one associated with the Mukhanov-Sasaki (MS) \cite{Mukhanov:1985rz,Sasaki:1986hm} equations. Furthermore, recent efforts to calculate the PPS by way of lattice simulation \cite{Caravano:2021pgc,Caravano:2021bfn,Inomata:2021tpx} look promising.\\ 

The rest of the paper is organized as follows. In the following section, we present the problem, our notations and our initial motivation for this study. In section \ref{sec:Methods_Analysis} we present the numerical methods we employ and the test cases we use. We also present the results and find the validity limits for the usual perturbative treatment. In the final section we summarize the results, present a `rule of thumb' for analytic validity, and discuss some of the consequences of these findings.\\

In this work we use the convention of $\hbar=c=M_{pl(\mathrm{ reduced})}=1$ and a mostly positive signature (-,+,+,+).
\section{The problem}
We address the case of an FRW metric, and a single field inflationary potential such that:
\begin{align}
    ds^2 = -dt^2 +a^2(t)d\Vec{x}^2,
\end{align}
and the action is
\begin{align}
    \mathcal{S}=\int d^4x \sqrt{-g} \left[\frac{-\mathcal{R}}{2}-\frac{\left(\partial_{\mu}\phi\right)^2}{2}-V(\phi)\right],
\end{align}
where $a(t)$ is the scale factor, $\mathcal{R}$ is the Ricci scalar, $g$ is the trace of the metric, $\phi$ is the inflaton field, and $V(\phi)$ is the inflationary potential.
~Since this type of scenario is the most extensively explored, studying this case both covers a large class of models and provides a proxy for different cases.\\

The PPS can be expressed as a power law of mode numbers:
\begin{align}
    P_{S}=A_{S}\left(\frac{k}{k_{0}}\right)^{n_s-1+\tfrac{\alpha}{2}\upsilon +\tfrac{\beta}{6}\upsilon^2 +\hdots},\label{Eq:PPS}
\end{align}
where $k_0$ is the pivot scale, $A_S$ is the amplitude at $k_0$ the pivot scale, $\upsilon=\ln\left(\frac{k}{k_0}\right)$, $n_s$ is the scalar spectral index, $\alpha$ is the index running and $\beta$ is the running of running. The pivot scale is chosen to minimize the correlation between $n_s$ and $\alpha$ \cite{Cortes:2007ak}, and is data-dependent. For instance, WMAP observations analyses were performed with $k_0=0.002\;Mpc^{-1}$ \cite{WMAP:2012fli}, whereas the Planck collaboration analyses mostly use $k_0=0.05\; Mpc^{-1}$ \cite{Planck:2018vyg}. The most commonly used relation between inflationary potential and PPS is the following Stewart-Lyth \cite{Stewart:1993bc} expression, and it's Lyth-Riotto \cite{Lyth:1998xn} variant and extension: 
\begin{align}
    n_s=&\simeq 1-6\varepsilon_{\scriptscriptstyle{V}}+2\eta_{\scriptscriptstyle{V}}\label{Eq:SL_ns}\\
    \nonumber & +2\left[\frac{\eta_{\scriptscriptstyle{V}}^2}{2} -\left(8b+1\right)\varepsilon_{\scriptscriptstyle{V}}\eta_{\scriptscriptstyle{V}} -\left(\frac{5}{3}-12b \right)\varepsilon^2_{\scriptscriptstyle{V}} +\left(b+1\right)\xi^2\right],
\end{align}
\begin{align}
    &\alpha \simeq -16\varepsilon_{\scriptscriptstyle{V}}\eta_{\scriptscriptstyle{V}} +24\varepsilon^2_{\scriptscriptstyle{V}} +2\xi^2,\label{Eq:LR_alpha}
\end{align}
with $b\simeq 0.73$, $\varepsilon_{\scriptscriptstyle{V}}=\frac{1}{2}\left(\frac{V'}{V}\right)^2$, $\eta_{\scriptscriptstyle{V}}=\frac{V''}{V}$,$\xi^2=\frac{V'V'''}{V^2}$, in which the notation $()'$ denotes a derivative W.R.T the inflaton field expressed as $\phi$.\\

Other attempts at recovering an analytical connection have been made (e.g. \cite{Dodelson:2001sh,Gong:2001he,Stewart:2001cd,Schwarz:2001vv}). Be that as it may, previous works \cite{Wang:1997cw} have pointed at a possible limit to the validity of these. Furthermore, some early numerical calculations \cite{Wolfson:2016vyx,Wolfson:2021utn} revealed glaring disparities between analytical predictions and numerical results, for some models, at a level that is discernible with our current CMB experiments.\\

We assume the validity of the analytical terms in Eqs. (\ref{Eq:SL_ns}, \ref{Eq:LR_alpha}) is mathematically sound to some limit. But we know of at least one case \cite{Ben-Dayan:2009fyj,Wolfson:2016vyx} in which all analytical attempts fail to reach consistent accuracy of better than $1\%$ required in the age of precision cosmology.\\
Thus, we expect to see a transition in accuracy when the field excursion $\Delta\phi\simeq 1$,  from high accuracy to low accuracy and maybe even as far as having no predictive value. We stress that all models evaluated are slow-roll models as all through the inflationary period the slow-roll parameters conform to $\epsilon_{\scriptscriptstyle{H}}\equiv\frac{-\dot{H}}{H^2}\ll 1$ and $\left|\delta_{\scriptscriptstyle{H}}\right|\equiv \left|\frac{\ddot{\phi}}{H\dot{\phi}}\right|\ll 1$. Models in which that was not the case were filtered out after calculating their background evolution.\\

\section{Numerical Methods and Analysis\label{sec:Methods_Analysis}}
\subsection{Methodology and Measures\label{sec:methodology}}
This section explains how we measure perturbations amplitude, and the response in the PPS. We stress that the analysis is made at the mathematical level, and the only physical constraint we demand is the slow-roll of the scalar field. \\
The amplitude of a perturbation over some baseline potential is given by an adjusted $L2$ functional norm:
\begin{align}
    A_{\mathrm{Pert}}=\frac{\sqrt{\int\left(V_{\mathrm{Pert}}-V_{\mathrm{PL}}\right)^2 d\phi}}{\int V_{\mathrm{PL}}d\phi }\label{eq:Pert_amp},
\end{align}
where $V_{\mathrm{PL}}$ is the power-law inflation potential, $V_{\mathrm{Pert}}$ is the perturbed potential and the integral is over the $\phi$-interval of interest. A perturbation over the baseline inflationary potential will typically yield a response in the PPS. When analysing response to a step-like feature and a Gaussian feature we look at the entire shape of the PPS. We do so to show that simple first order approximations of the PPS may be insufficient, and higher moments are called for. However, it is fairly well established that sharp features in the PPS are not likely, as they would leave other signals, such as spectral distortions \cite{Lucca:2019rxf, Kite:2020uix}.\\ 

Next, we perturb over two classes of base-potentials. First we look at LFMs as a generalization of the exponential potential that yields a power-law inflation. Then we look at SFMs as a generalization of a monomial potential of the form $V(\phi)=1-a_p\phi^p$. In both cases we measure the departure of the spectral index and index running as predicted by analytical terms, from numerical results. For this we use the measure of asymmetry:
\begin{align}
    Asym(p_1,p_2)=200\left(\frac{p_1-p_2}{p_1+p_2}\right),\label{Eq:Asym}
\end{align}
which is the difference divided by the mean, expressed as percents. For instance, the asymmetry between an analytically derived $n_s^{\textrm A}$ and a numerically calculated $n_s^{\textrm N}$ one is given by:
\begin{align}
    Asym(n_s^{
    \textrm A},n_s^{\textrm N})=200\left(\frac{n_s^{
    \textrm A} -n_s^{\textrm N}}{n_s^{
    \textrm A} +n_s^{\textrm N}}\right),
\end{align}
with the result expressed in percents. One of the pitfalls of this analysis is for values close to zero that may change signs, thus reducing the denominator. For this reason when quantifying the asymmetry of minute values, such as $\alpha$, we use $n_s$ in the denominator, i.e.:
\begin{align}
    Asym(\alpha^{
    \textrm A},\alpha^{\textrm N})=200\left(\frac{\alpha^{
    \textrm A} -\alpha^{\textrm N}}{n_s^{
    \textrm A} +n_s^{\textrm N}}\right).
\end{align}
This has the additional merit of easily assessing whether indeed $\alpha\sim \left(1-n_s\right)^2$.\\

In the case of LFMs and SFMs analysis, we do not simulate features. Rather, we perturb over the coefficients in a Taylor expansion of the base potential. This perturbed potential has to pass two tests. First, a design requirement is that:
\begin{align}
    60\gtrsim\int_{\phi_{\mathrm{start}}}^{\phi_{\mathrm end}}\frac{-V}{V'}.
\end{align}
This ensures that we always look at well behaved models. Secondly, as stated before we only analyse models which, after background evolution, yield slow-roll inflation. Thus there are no sharp features in the perturbed physically-oriented classes.
\subsection{Numerical Process}
We use the previously employed INflationary potential Simulator and ANalysis Engine (INSANE\footnote{Code available at \url{https://github.com/beastraban/INSANE}}) package \cite{Wolfson:2021utn}, with additional artificial intelligence elements, developed for the study of small field models.\\

The INSANE package's input is either a symbolic or numeric representation of an inflationary potential. It solves the background Friedmann equations, retrieving the cosmological quantities required to build the associated MS equations \cite{Mukhanov:1985rz,Sasaki:1986hm} most easily written in conformal time:
\begin{align}
    \partial_{\tau\tau}U_{k}+\left(k^2-\frac{ \partial_{\tau\tau}Z}{Z}\right)U_k=0,
\end{align}
where the pump field $Z$ is given by $Z=\tfrac{a\dot{\phi}}{H}$ and $U_k$ are the perturbation wave function eigenmodes. The code solves this equation for a large number of these $k$-modes, typically 120. It then analyzes the resulting PPS for the scalar index, its running and possibly higher moments. Finally it saves all corresponding quantities to file, including a paired graphs for potential and PPS.\\

The assignment of physical k-numbers is done after the background geometry has been derived. When we evaluate physical models to recover the spectral index and other observables, we find the point in the background evolution where the scale factor satisfies $\log(a_0)-\log(a(t))=N$, where $a_0$ is the scale factor at the end of inflation and $N$ is a predefined number of efolds. The inflaton field value at that point is called $\phi_{\mathrm{CMB}}$.\\

This work, however, discusses the correlation between analytic and numerical results. The analytic results are always defined at the CMB point. Thus we always define $\phi_{\mathrm{CMB}}=0$ and $k_0$ as the corresponding mode number. This is done to promote a clear Taylor-series representation around $\log(k/k_0)=0$ and $\phi=0$. If required we can then shift the horizontal axis to match the computational scale $k$ to the physical scale $\Tilde{k}$ and recover the power spectrum observables at different physical pivot scales. This further complicates the process of model building, as the physical pivot scale usually differs from the scale at which initial conditions for the inflaton field are set. This just emphasizes the point, since if the analytical approximations do not match the numerical response at a definitionally identical point, there is little chance of proper predictive power at a different scale.\\

\begin{figure}[!h]
    \centering
    \includegraphics[width=0.9\textwidth]{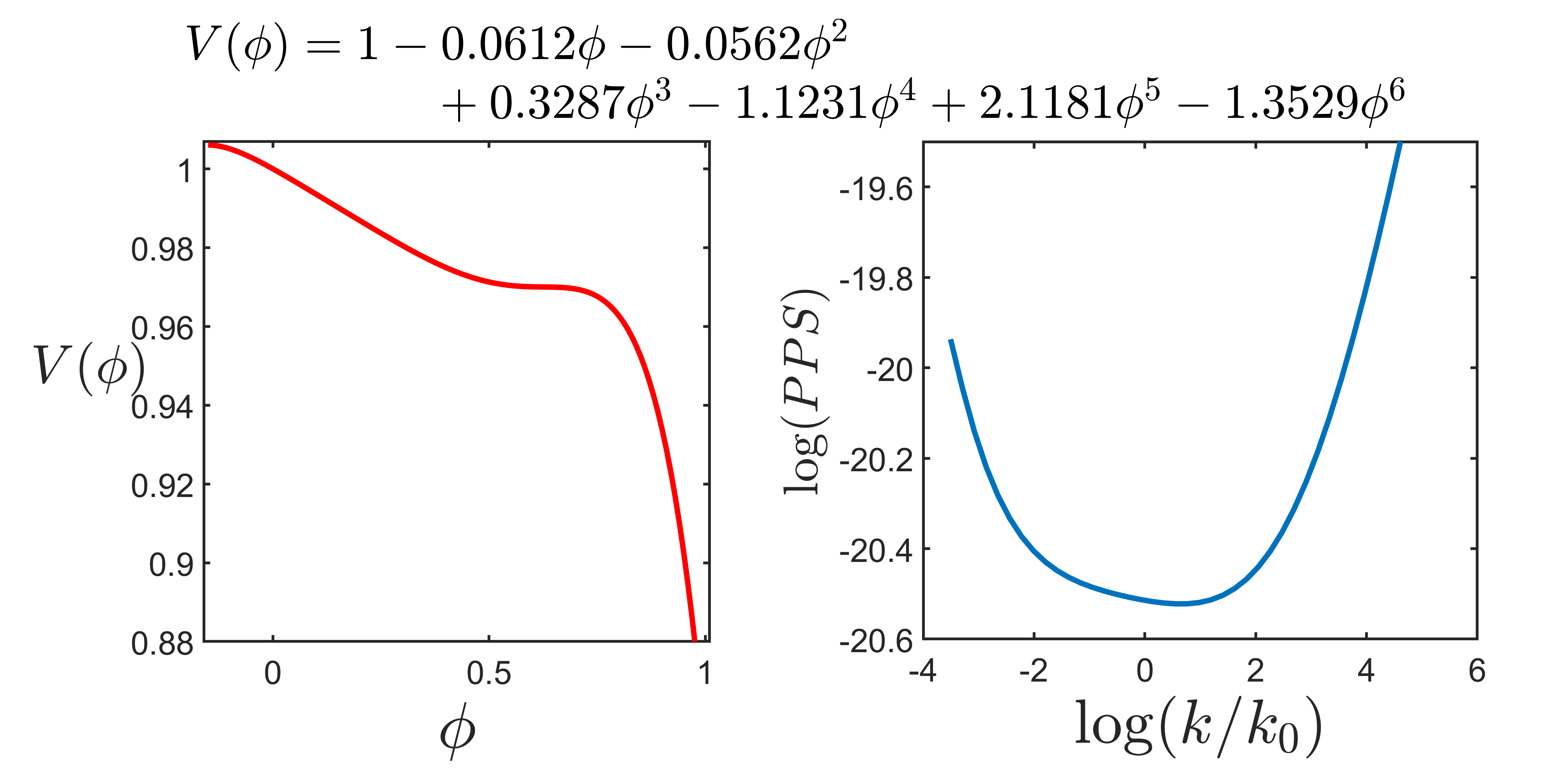}
\caption{A typical output of the INSANE package. A potential-PPS pair of graphs, as well as a file containing the $\log(k)$, $\log(PPS)$ data, and additional information.}
    \label{fig:INSANE_TYPICAL_RESULTS}
\end{figure}

A characteristic output of the INSANE package for some arbitrary polynomial model, is given in Figure~\ref{fig:INSANE_TYPICAL_RESULTS}. \\

\subsection{Step Function Approximations}
\begin{figure}[!h]
    \centering
    \includegraphics[width=0.8\textwidth]{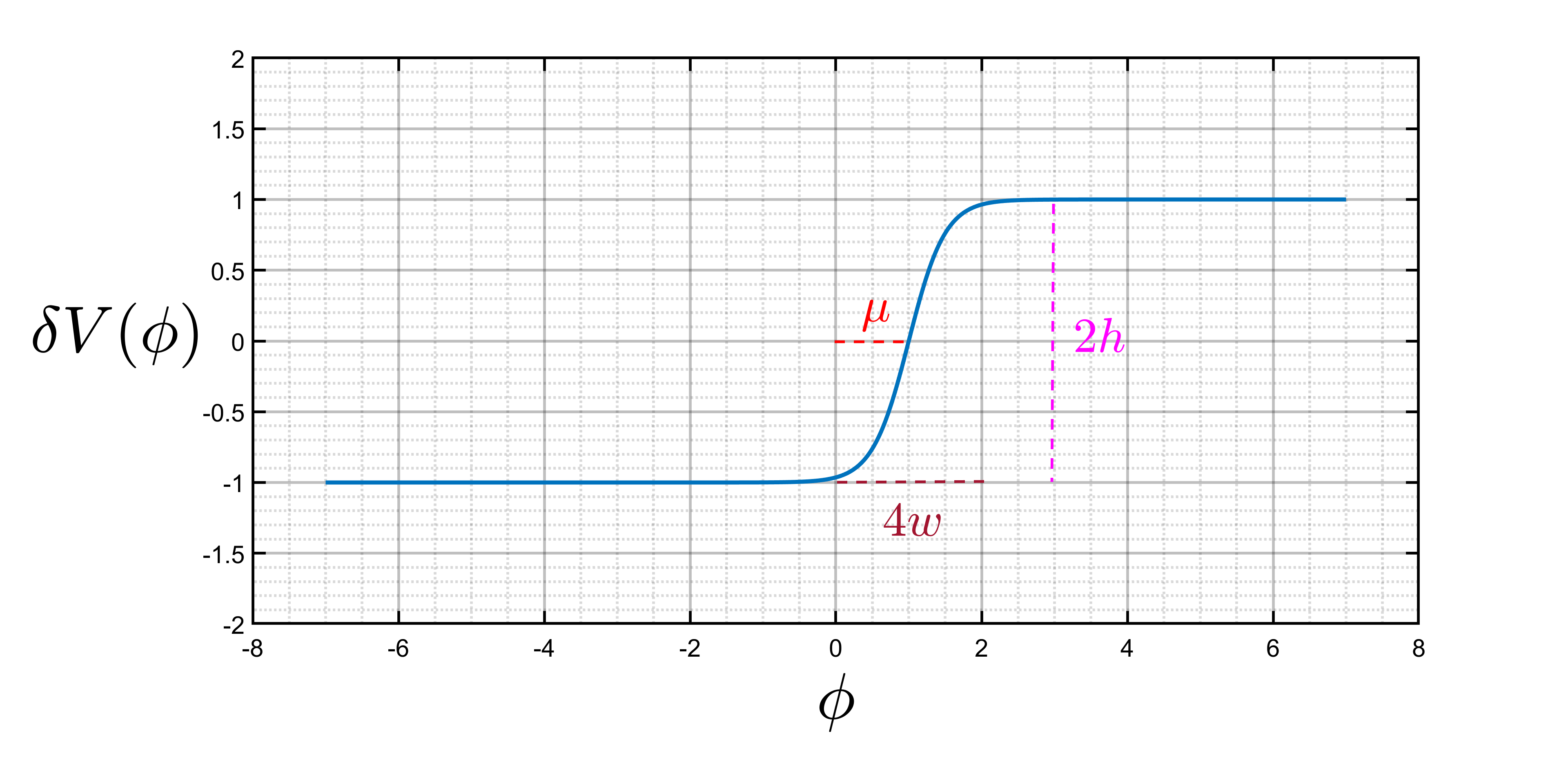}\\
    \includegraphics[width=0.85\textwidth]{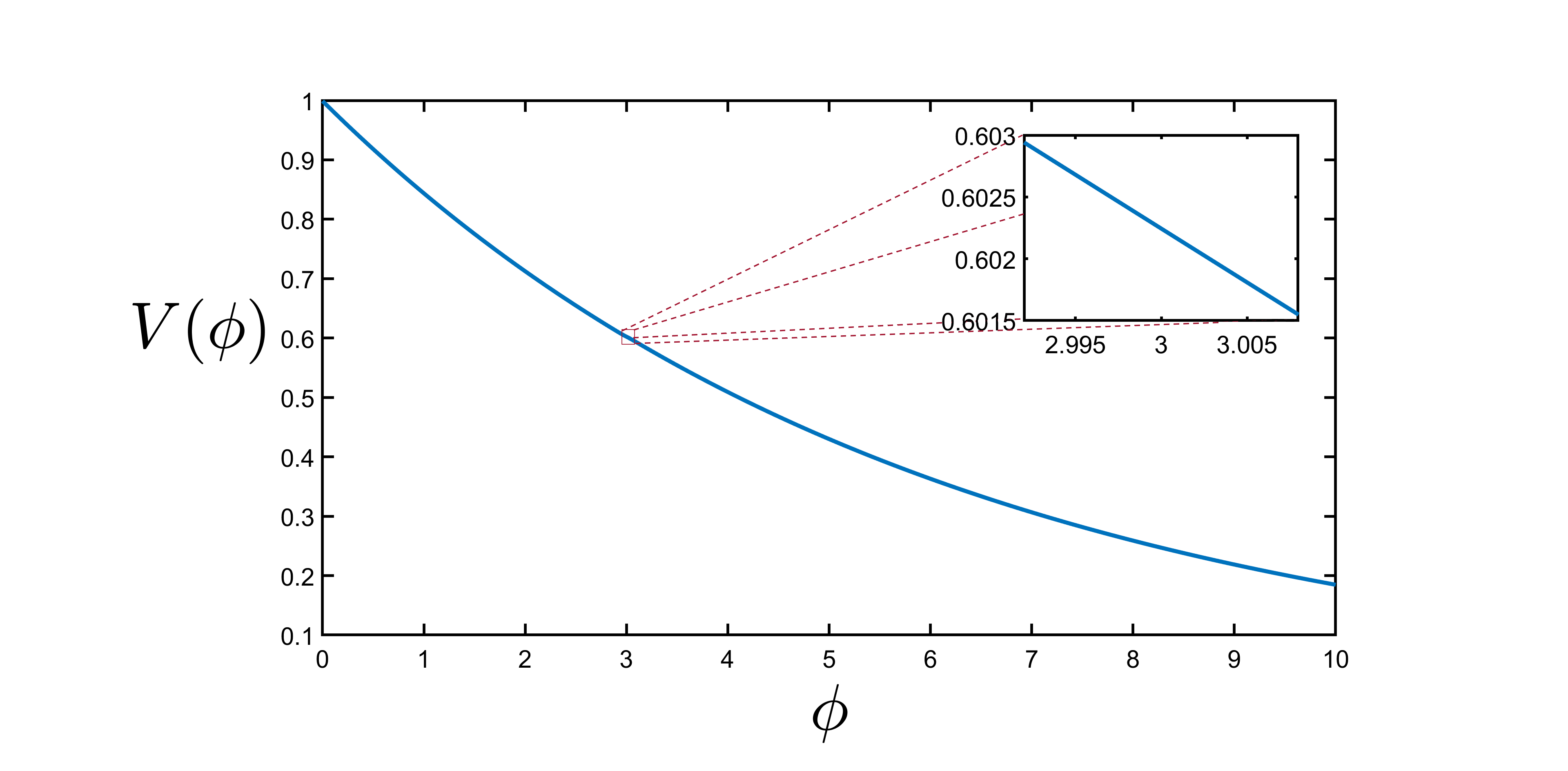}
    \caption{A hyperbolic tangent function added to a power-law inflationary potential. In the upper panel we see the characteristics of the hyperbolic tangent function. The lower panel is the most pronounced function we add over the base function: $V=\left[V_0 \exp\left(-\sqrt{\tfrac{2\phi^2}{70}}\right)\right]\left(1+0.001\tanh\left( \frac{\phi-3}{0.04}\right)\right)$. We see that even the maximal disturbance is hard to discern.}
    \label{fig:tanh_example}
\end{figure}
We systematically study features in the inflationary potential, to quantify the effect a departure from power-law inflation might have on the power spectrum. 
In the case of a step function, as well as in the next subsection dealing with Gaussian features (\ref{sec:Gaussian}), the power-law inflationary potential $V(\phi)=V_0 \exp\left(-\sqrt{2\phi^2/P}\right)$ with $P=70$ is used as a baseline. This is done due to the case of power-law inflation being the only fully analytical solution for the background and perturbation equations. In the power-law case, the eigenfunctions of the MS equation are given by a set of Bessel functions. After applying initial conditions we are left with a set of Hankel functions, which we then assess at a certain late time ($
\tau\rightarrow 0_{\scriptscriptstyle{\mathbb{-}}}$), to recover the $PPS(k)$ function. The power-law case is well documented, and yields a constant spectral index of $n_s=1-\frac{2}{P}$.\\
\begin{figure}
    \centering
    \includegraphics[width=1\textwidth]{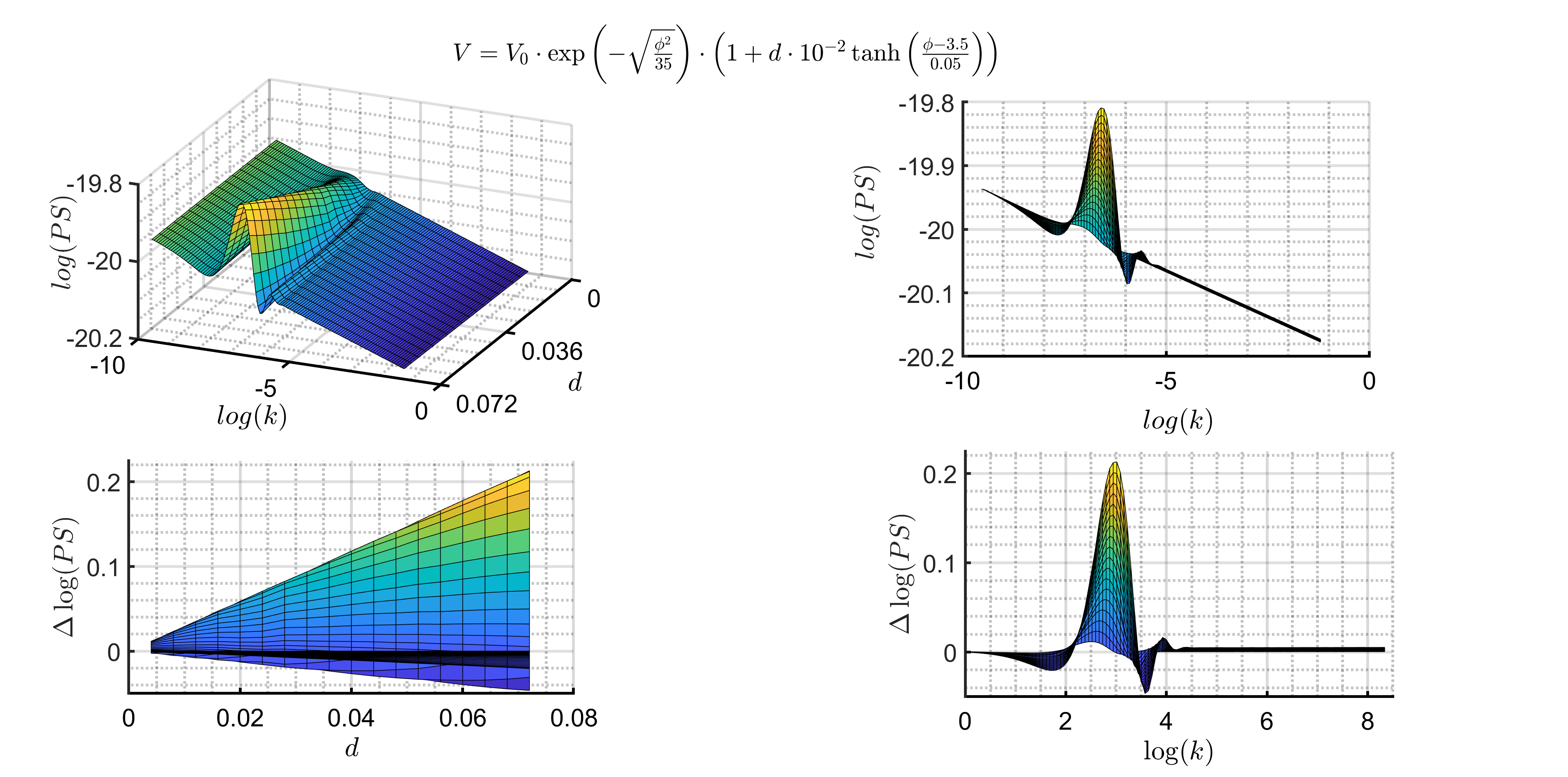}
    \caption{The power spectra surface for different heights of step-like perturbation with width of $\Delta\phi=4\times 0.05=0.2$ (left panels). And a stacked view of power spectra (right panels). We remove the power-law power spectrum from the result to better see the PPS behavior (lower panels).}
    \label{fig:step_const_d}
\end{figure}
We use the hyperbolic tangent function to simulate a step function, similar to what was done in \cite{Adams:2001vc}. Specifically, we use the following functional form:
\begin{align}
    V(\phi)=\left[V_0 \exp\left(-\sqrt{\tfrac{2\phi^2}{70}}\right)\right]\left(1+h\tanh\left( \frac{\phi-\mu}{w}\right)\right),
\end{align}
in which $0<h\ll 1$ is the height of the step, $\mu$ is the location of the step on the $\phi$ axis and $w$ correlates to the width of the step. Figure~\ref{fig:tanh_example} demonstrates the features of a hyperbolic tangent function, and an addition of the sharpest step-like function we use to the basic power-law inflationary potential. We can see that even in the most pronounced perturbation, the difference is negligible, and as such should conform to the notion of small perturbative corrections to the power-law inflation's fixed $n_s$.

We run complete numerical simulations for perturbations of height $0-0.1\%$ over the base potential with $V_0=1$, and of tangent width between $\Delta\phi=0.04-0.6$. The location of the perturbation was consistent through all simulations. However, due to changes in $\dot{\phi}$ caused by the potential perturbation, the PPS response to the feature may begin in slightly different scales $k$. Thus, to promote meaningful comparison, we shifted the resulting power spectra such that the feature response start at the same mode number $k_0=1$. \\

\begin{figure}[!h]
    \centering
    \includegraphics[width=1\textwidth]{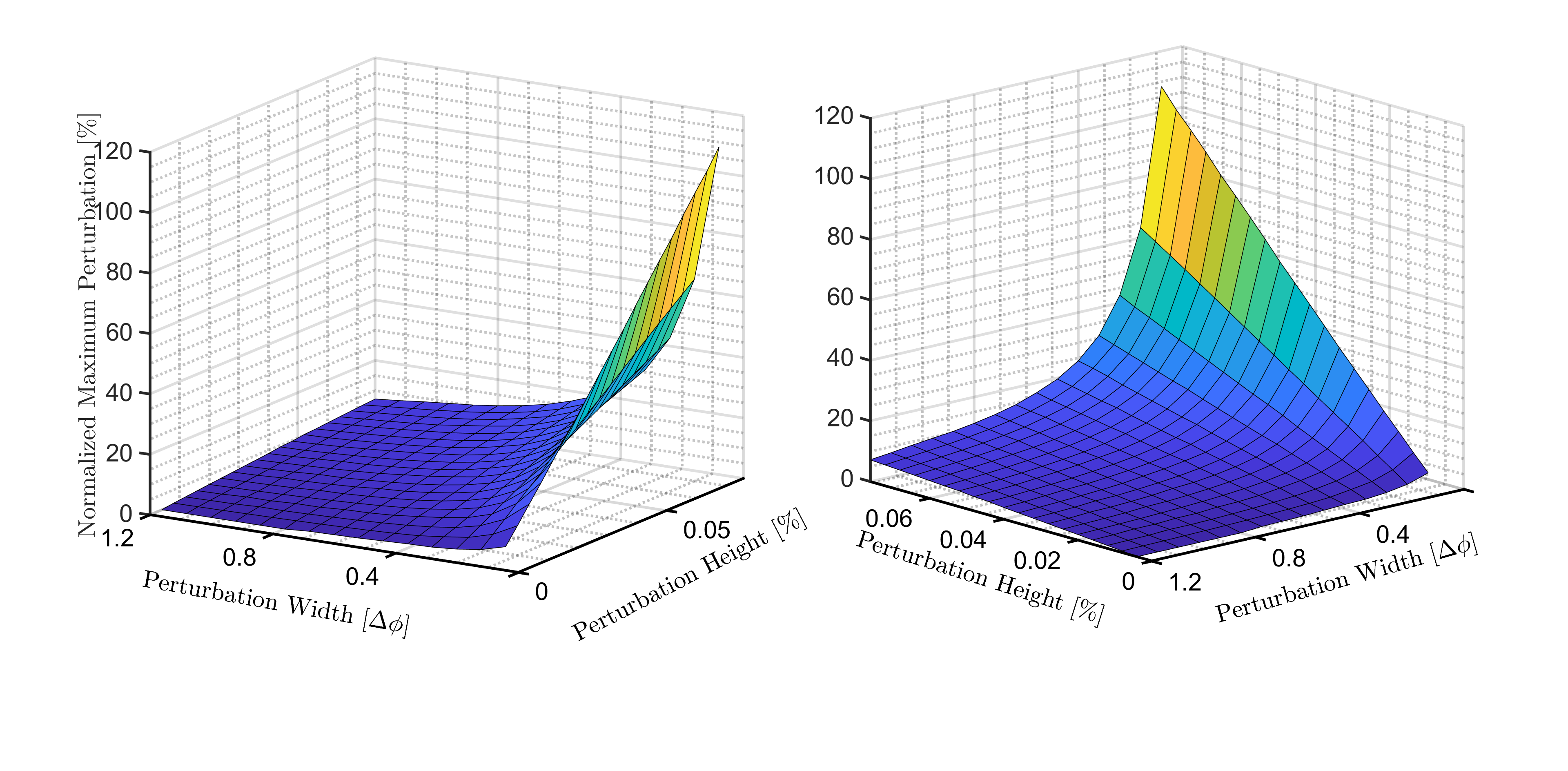}
    \caption{The response landscape to step-like perturbations. We see that in the lower end of perturbation  significance, the spectra are well behaved whereas for a feature of width smaller than $\Delta\phi \sim 0.4$, even a mild perturbation yields a significant departure from the pure power-law spectrum. }
    \label{fig:step_pert}
\end{figure}

Quantifying the perturbative response involves several metrics as discussed in \ref{sec:methodology}.  As shown in Figure~\ref{fig:step_const_d}, first we look at the resulting power spectra, and remove the base power-law spectrum. We then take the perturbed part and shift the baseline such that $0$ perturbation is assigned the base value $1$. The first metric we use is the maximum height of the response with relation to the base value. We record these for perturbations of height $h\in\left[0,0.72\right]\%$ and width $4w\in \left[0.04,1.2\right]$ and find the maximum height of the perturbation in terms of percents as shown in Figure~\ref{fig:step_pert}.  

When using this metric we try to fit the entire perturbed spectrum with a first degree polynomial of the form $\log(PPS)=C+(n_s^{(1)}-1)\log(k)$, with $C$ being some constant, and the superscript denotes the degree of polynomial used in fitting. we then plot the $n_s$ asymmetry between the power-law spectrum and the perturbed spectrum. Since the fit is taken around $\log(k)=0$ the theoretic fit should be the power-law one. Regardless, we always re-calculate the analytical scalar index according to eq.~\ref{Eq:SL_ns} at $\phi_{\mathrm{CMB}}$.
\begin{figure}[!h]
    \centering
    \includegraphics[width=0.8\textwidth]{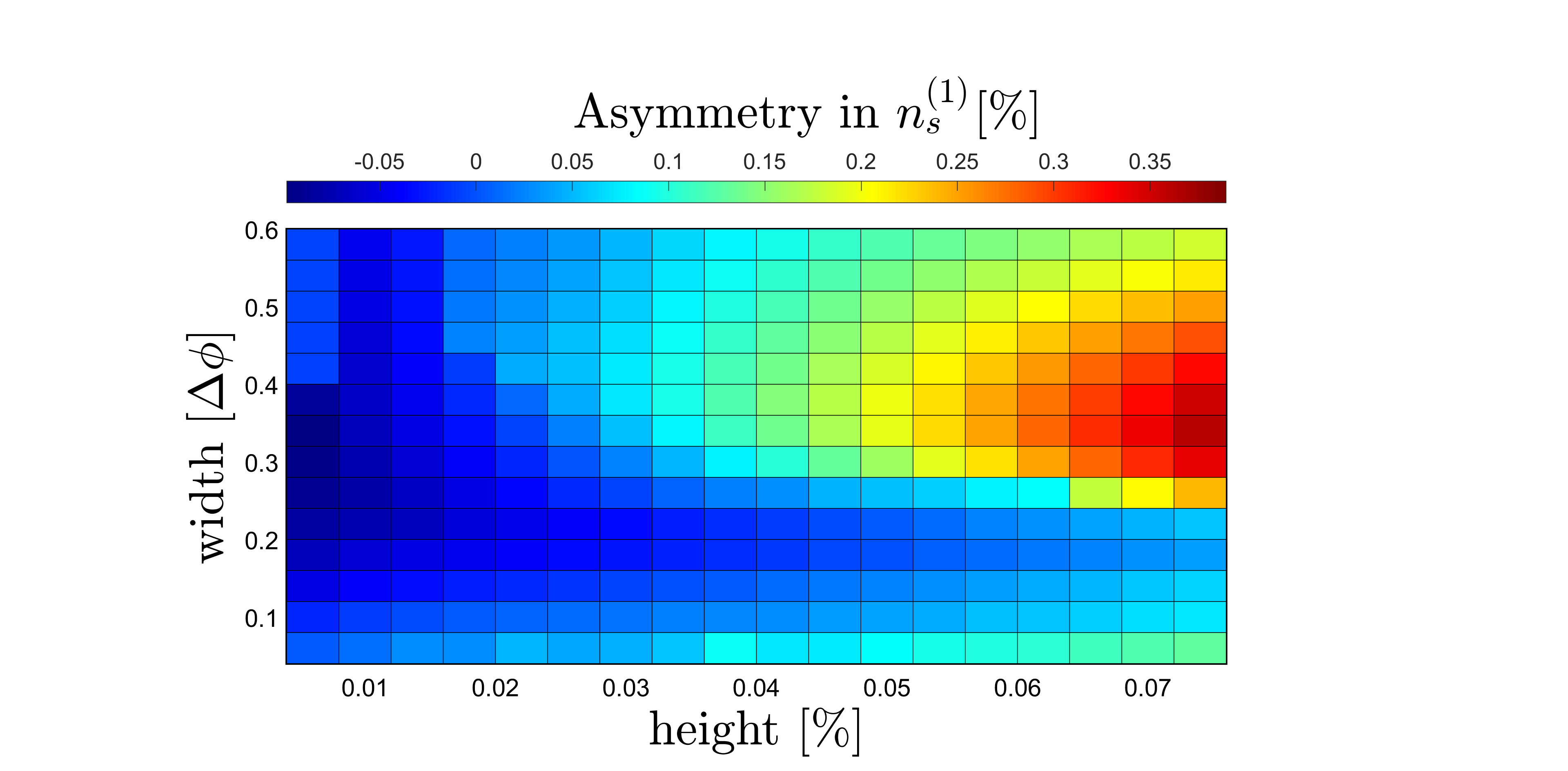}
    \caption{The maximum departure of $n_s^{(1)}$ from theory is $\sim 0.35 \%$ (see top colorbar), where the superscript denotes the degree of fitting polynomial. This is misleading due to the first order polynomial fit missing the feature. Also note that the maximum departure is around the highest and widest features, contrary to expectations.}
    \label{fig:ns1_step}
\end{figure}

Figure~\ref{fig:ns1_step} demonstrates that the departure from theory is at most $0.35\%$ when fitting to first degree. This is due to the perturbation being approximately evenly distributed above and below the baseline, thus affecting the vertical axis intercept $C$ but not the slope. So the first degree polynomial fits the spectrum sufficiently well but does not reflect the feature.

\begin{figure}[!h]
    \centering
    \includegraphics[width=0.8\textwidth]{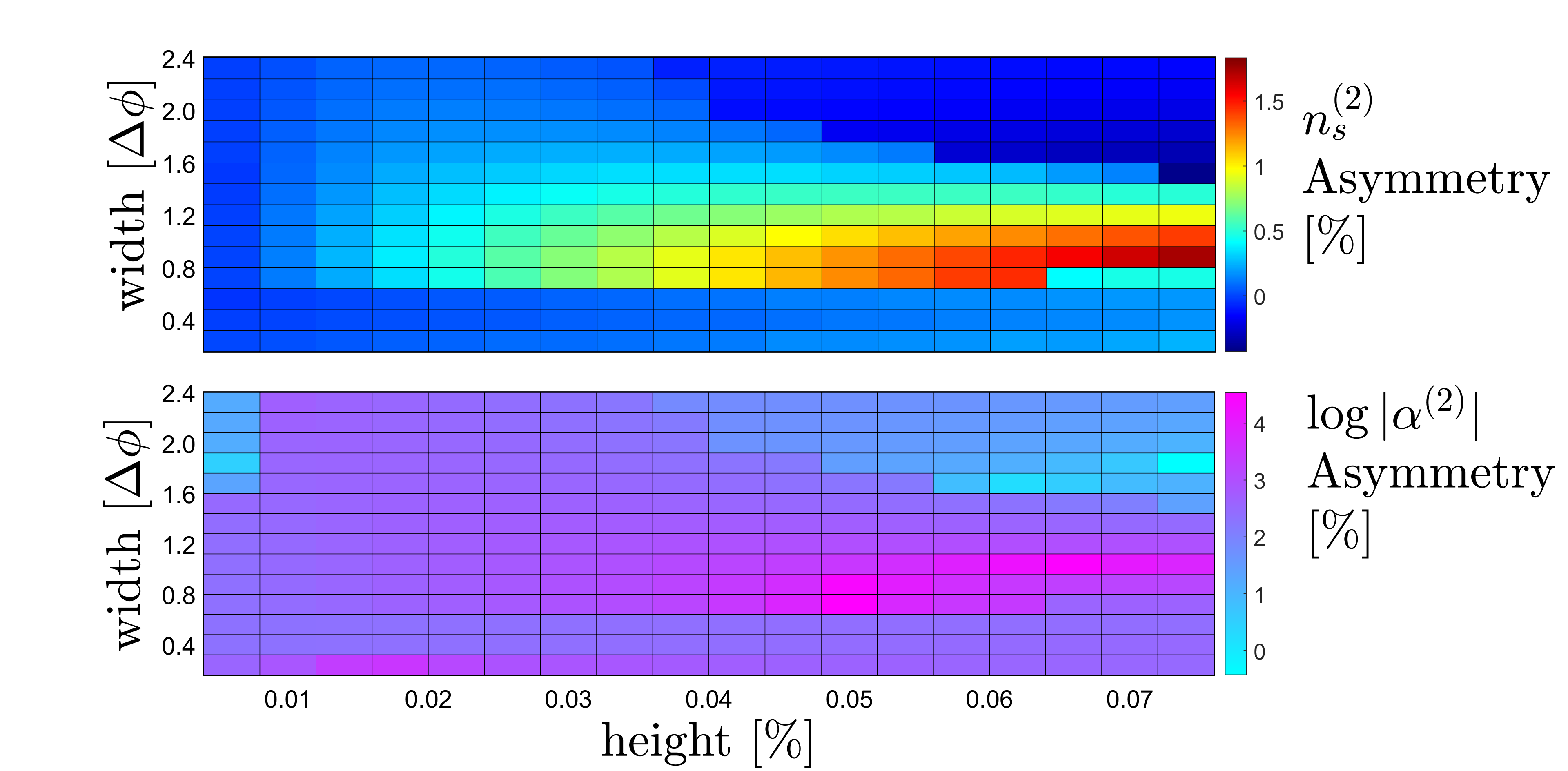}
    \caption{Results of fitting the perturbed spectra with a quadratic function. We see the $n_s^{(2)}$ asymmetry is sharper, and both $n_s^{(2)}$ and $\alpha^{(2)}$ asymmetries peak around $4w\sim 1.2$. Note that the asymmetry in $\alpha^{(2)}$ is on a logarithmic (base 10) scale. Additionally we see that when the potential feature is sharp enough, fitting the spectrum with a quadratic function cannot well describe the PPS response.}
    \label{fig:ns_nrun_step}
\end{figure}
When fitting with a higher degree polynomial, we better capture the presence of a feature, but relinquish our ability to accurately predict $n_s$. This is shown in Figure~\ref{fig:ns_nrun_step}, where we fit the power spectra with a quadratic function, to extract the running of the scalar index $\alpha^{(2)}$. 
\begin{figure}[!h]
    \centering
    \includegraphics[width=0.8\textwidth]{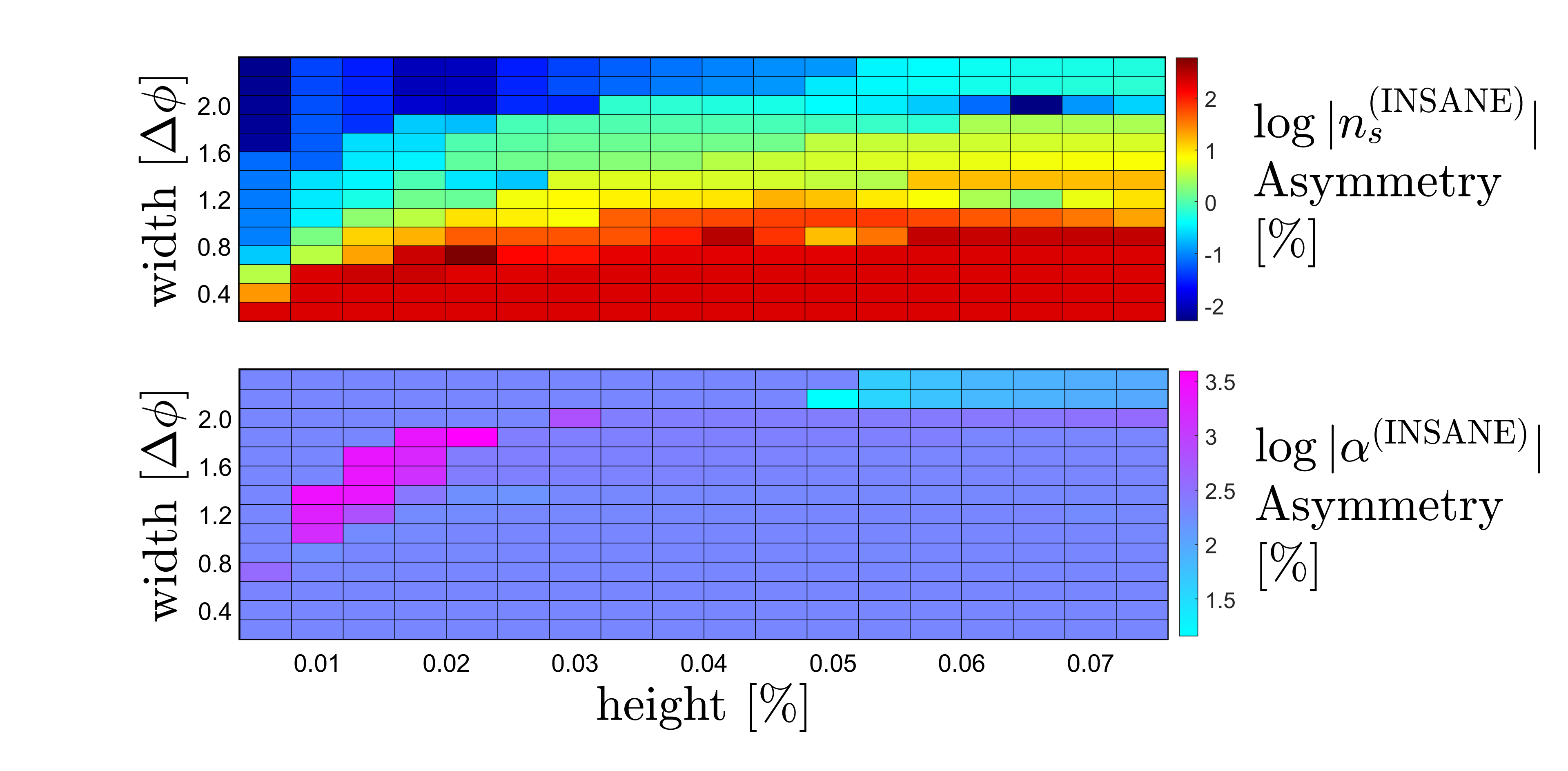}
    \caption{A polynomial approximation of the power spectra to cumulative error better than $10^{-4}$, shows the expected behavior in $n_s$ asymmetry. We also see that the $\alpha$ asymmetry is mostly around $100\%$. }
    \label{fig:Ultimate fitting}
\end{figure}
This suggests that to capture the entire form of the spectra higher moments are needed. Thus, we fit the spectra with the minimal polynomial series that complies with the accumulated error threshold set to $10^{-4}$. The cumulative error is given by:
\begin{gather}
    \delta_{\mathrm{error}} = \sqrt{\sum_{i=1}^N \left(f_{i}-g_{i}\right)^2},
\end{gather}
with $f$ being the simulated PPS, and $g$ its polynomial fit.
Figure~\ref{fig:Ultimate fitting} shows that where the width of the feature is less than $\Delta\phi\simeq 0.8$, for all but extremely small perturbations the analytically derived scalar index $n_s$ is off by $\mathcal{O}(10\sim 100)\%$.
We note that the minimal asymmetry in $\alpha$ is of order $\sim 30\%$. 
\subsection{Gaussian Feature}\label{sec:Gaussian}
\begin{figure}
    \centering
    \includegraphics[width=0.85\textwidth]{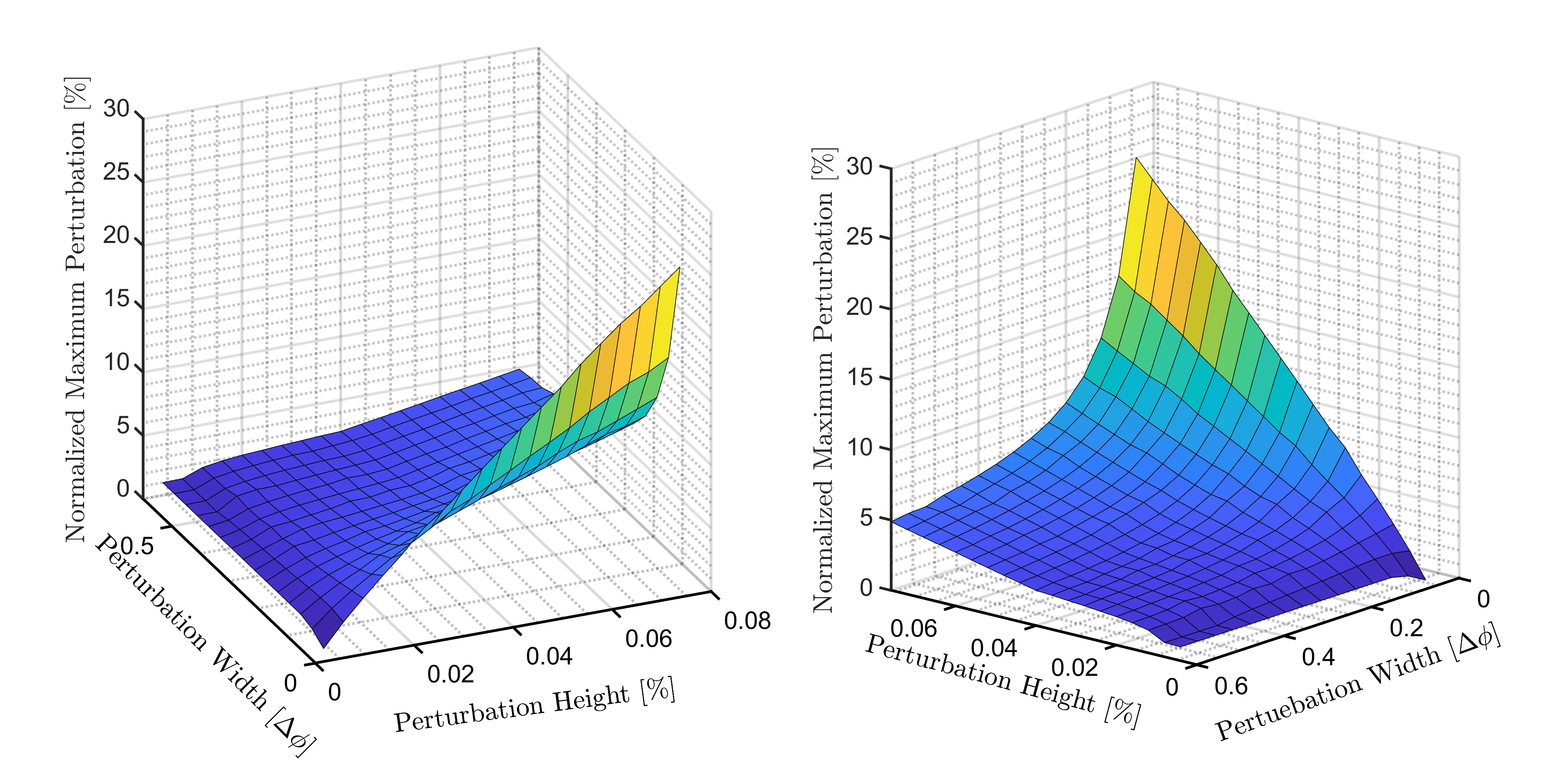}
    \caption{The maximal departure of the perturbed power spectra from the power-law base shows a disproportional effect of the Gaussian perturbation on power spectrum response.}
    \label{fig:Maximal_Gaussian}
\end{figure}

\begin{figure}
    \centering
    \includegraphics[width=0.85\textwidth]{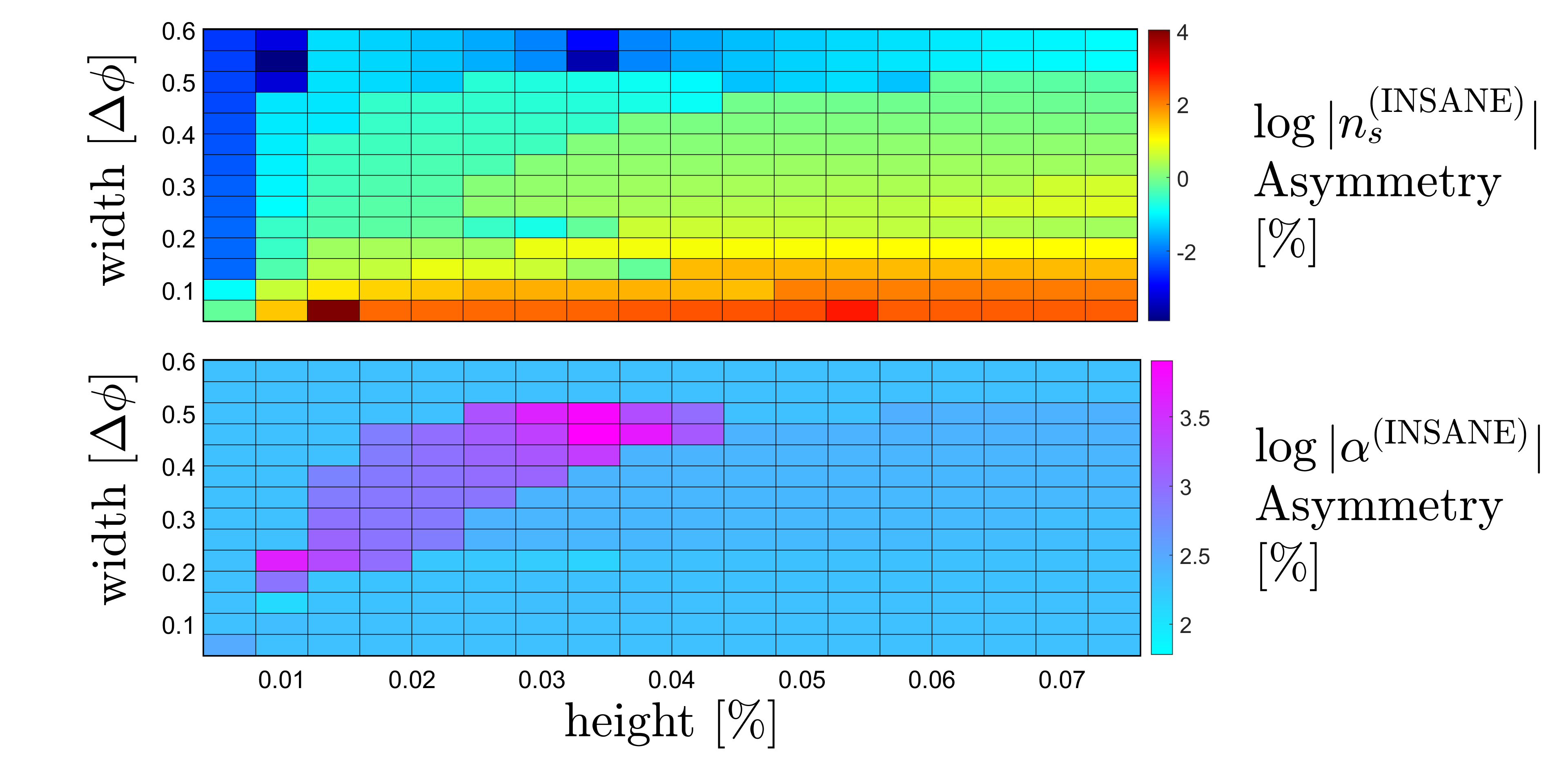}
    \caption{An analysis of the asymmetry between calculated analytical power spectra and their numerical counterparts shows a marked deviation. As expected for wider and smaller perturbations we see little to no effect with an asymmetry of order $<1\%$. However when the feature becomes sharper, we have a usual asymmetry upwards of $10\%$ and up to more than $100\sim1000\%$. The minimal asymmetry in $\alpha$ is around $70\%$}
    \label{fig:GAUSSIAN_ULTIMATE}
\end{figure}

The same process as the one in the previous section is applied to a Gaussian feature over a power-law inflation base potential. The dimensions of the Gaussian feature are height of $h\in\left[0.004,0.076\right]\%$, and width $2\sigma^2\in\left[0.04,0.6\right]$. 
\begin{figure}[!h]
    \centering
    \includegraphics[width=0.8\textwidth]{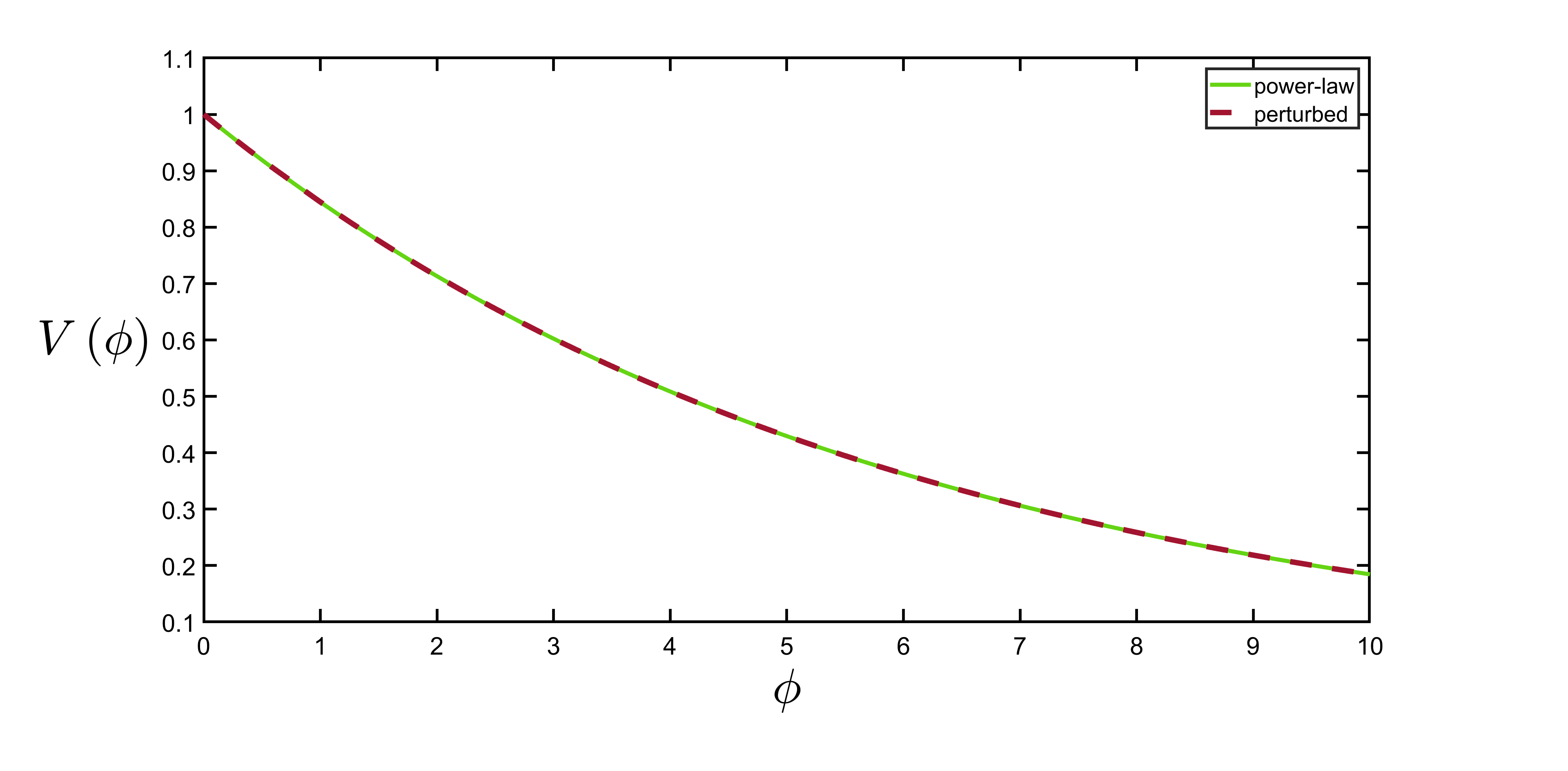}
    \caption{A comparison between the power-law potential and the potential with a Gaussian feature of relative height of $0.1\%$ and width of $\sigma^2=0.04$. This is a sharper feature than the maximal perturbation we use, but the perturbation is indiscernible.}
    \label{fig:Gaussian}
\end{figure}
The results for a Gaussian-perturbed potential reinforce our notion of perturbative treatment failing as features become sharper. As seen in Figure~\ref{fig:Gaussian}, the feature added to the overall power-law potential is very small compared to the overall potential scale. However, the departure from the power-law spectrum is disproportionally pronounced as Figure~\ref{fig:Maximal_Gaussian} shows. Comparing the results from a minimal polynomial with negligible error, to the analytical term by way of asymmetry (see Figure~\ref{fig:GAUSSIAN_ULTIMATE}) yields a usual asymmetry in $n_s$ of order $10\%$, and sometimes up to $1000\%$ or more. Moreover, the minimal asymmetry in $\alpha$ is around $70\%$.
\subsection{Corollary}
The prospects of a feature in the inflationary potential are not new. A presence of such a feature may yield a feature in the PPS. Thus, it has physical repercussions, for instance it may influence the effective number of neutrino degrees of freedom as shown in \cite{Ben-Dayan:2019gll,Meerburg:2015zua}. \\

It follows from Figure~\ref{fig:step_const_d} that where the feature in the inflationary potential becomes sharper, the PPS response becomes more pronounced. While the prospects of a sharp feature in the PPS were mostly ruled out, ultimately, even for slight features in the inflationary potential, the response may be ill-defined in terms of the usual power series representation for the PPS as in eq.~\ref{Eq:PPS}. \\

Thus, it would be beneficial to consider using the full form of the PPS in likelihood analyses. This is contrary to the common practice of considering only the linear PPS or at the most the quadratic case. However, as stated before, these possible features cannot be too pronounced. Such pronounced features would have created other signatures of cosmic observables, which we do not detect.\\

\subsection{Perturbative Treatment of Physical Models}
Supplying a rule of thumb for the validity of simple perturbative treatment requires a simple, testable heuristic criterion for evaluating the perturbation amplitude. While some heuristics may be more accurate, we aim for simplicity of use. In the following we justify the test introduced in \eqref{eq:Pert_amp}.
\subsubsection{Large Field Models}
Since the purely analytical case is the power-law inflation, we use the power-law inflation potential as the baseline. \\
\begin{figure}[!h]
    \centering
    \includegraphics[width=0.85\textwidth]{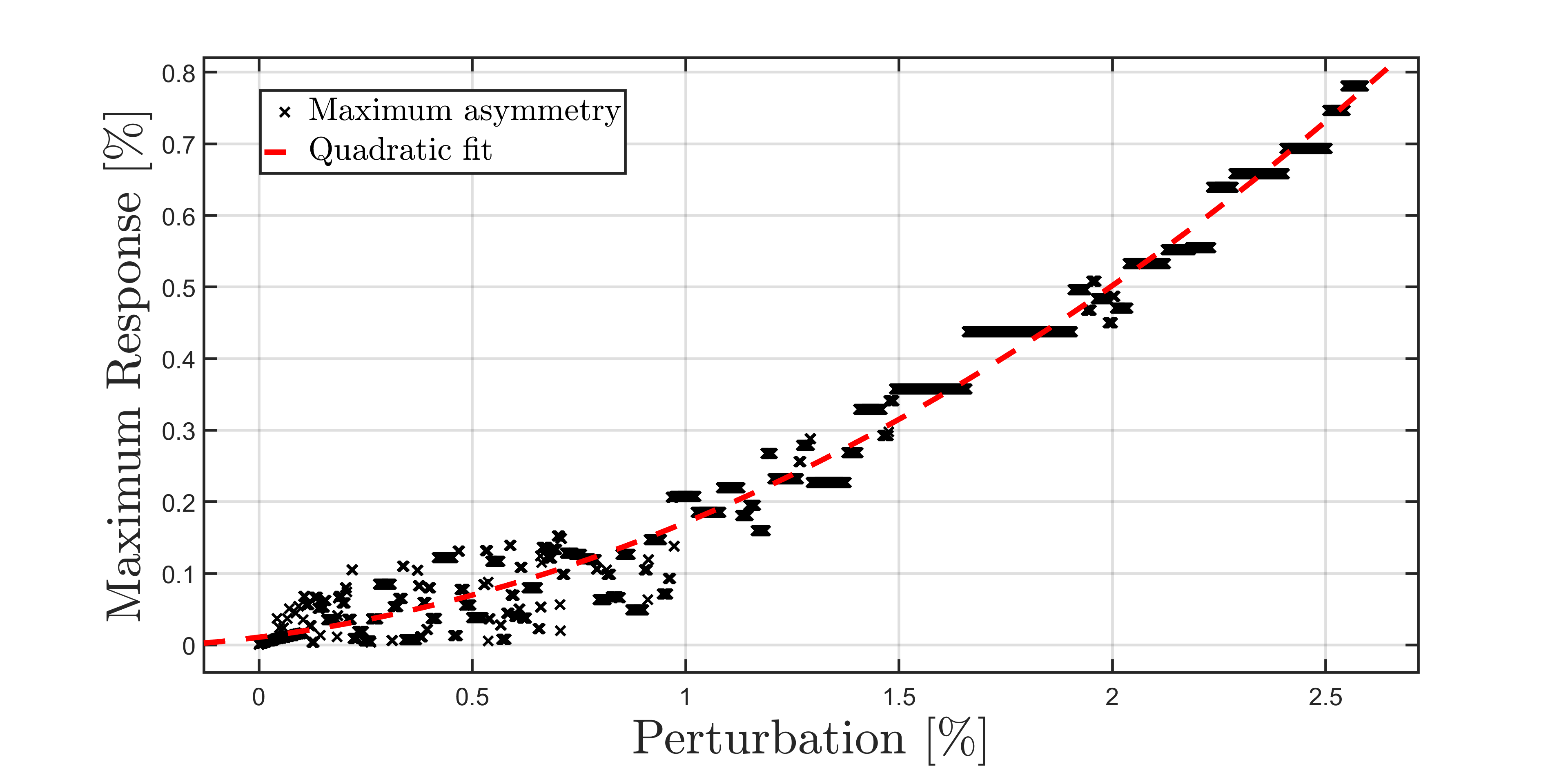}
    \caption{We probe the asymmetry response to perturbations over a power-law inflation potential as a stand-in for a generic large field model. We see that around the $3\%$ mark, the response exceeds the $1\%$ asymmetry limit required in the era of precision cosmology.}
    \label{fig:Large-Field-asym}
\end{figure}
The process of generating perturbed potentials is the following. We set the CMB point at $\phi_{\mathrm{CMB}}=0$, and Taylor expand around it to the $8th$ degree. We then set the maximum perturbation to $75\%$, and each coefficient is then multiplied by $1+ \Delta_i$, where $\Delta_i$ is randomly set for each coefficient separately such that $\Delta_i\in[-0.75,0.75]$. However, the amplitude of perturbation is assessed using an adjusted $L_2$ norm \eqref{eq:Pert_amp}.
 The response to the perturbation is measured by the scalar index' $n_s$ asymmetry (eq. \ref{Eq:Asym}) from the usual Lyth-Riotto term (eq.~\ref{Eq:SL_ns}) as calculated at $\phi=\phi_{\mathrm{CMB}}$. The inflaton coordinate at the CMB point should be $0$ by construction. However, numerically it is taken as the closest value to zero, of sampled $\phi$, which is of order $10^{-5}\sim 10^{-6}$.  We bin the asymmetry responses by perturbation amplitude, taking the maximum asymmetry value for each bin. We see that the maximal asymmetry is sufficiently fitted by a quadratic function. Furthermore, we deduce that the $n_s$ asymmetry should be below the $1\%$ mark for perturbations up to around $3\%$, which is apparent by inspecting Figure~\ref{fig:Large-Field-asym}. 

\subsubsection{Small Field Models}
For the small-field model case, we find ourselves in the proverbial pickle. There is no purely analytical small-field baseline to perturb over. Fortunately, in models of the form:
\begin{align}
    V_p(\phi)=1-a_p\phi^p, \label{eq:SFMpot}
\end{align}
which with $a_p>0$ can be viewed as an extension of the hilltop model, we can use a large $p$. Thus, we can ensure that in the first $\sim 8$ efolds of inflation the potential displays a fairly fixed slope. This translates to a sufficiently small time variability of slow-roll parameters that ensures the correctness of the perturbative treatment. However, the physically interesting models are those who display the observed scalar index. Furthermore, if inflation is to be examined by its GW signature, we require an appreciable tensor-to-scalar ratio $r$. These two considerations, are hard to reconcile with the requirement of $\Delta\phi\simeq 1$. This is because such a large $\varepsilon$ requires $\Delta\phi\simeq 0.35$ to generate the first $\sim 8$ efolds detected in the CMB. As was shown in \cite{Ben-Dayan:2009fyj,Choudhury:2014kma} , these models present an appreciable index running, and a manifestly time varying slow-roll parameters.\\

Since the coefficients in \eqref{eq:SFMpot} for $\phi^k \;, k<p$ are all set to $0$, there is no natural scale for a perturbative treatment of these coefficients. We thus use the following recipe. We shift $\phi\rightarrow \phi-0.5$, recover the coefficients, perturb over those, reset $\phi\rightarrow\phi+0.5$, and normalize the perturbed potential such that $V_{\mathrm{Pert}}\big|_{\phi=0}=1$. \\
\begin{figure}[!h]
    \centering
    \includegraphics[width=0.85\textwidth]{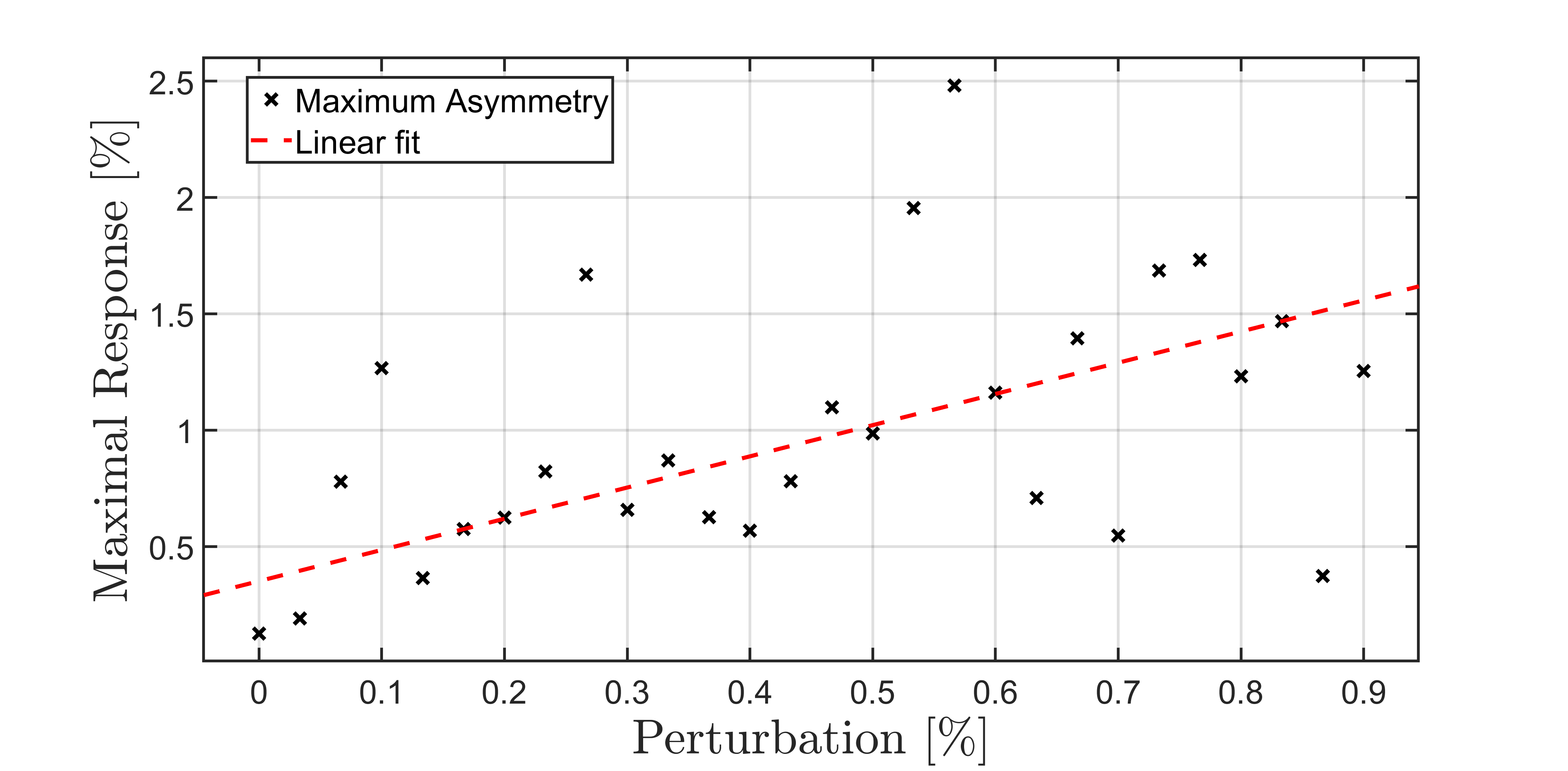}
    \caption{Examining the spectral response to perturbed SFMs we find it erratic. At a $0.1\%$ perturbation we see an asymmetry response of over $1\%$. Plotting the trend we find that already for $0.5\%$ perturbations we can expect a consistent $1\%<$ asymmetry in response. The fluctuations are due to the limited number of viable models produced.}
    \label{fig:SMF_fullAssymetry}
\end{figure}

We apply this process for $p=4$. First we calculate $a_4=1/(2\sqrt{2}+1)$ by requiring $\varepsilon_{\scriptscriptstyle{V}}=1$ at $\phi_{\mathrm{end}}=1$. We do this to set the small field excursion during inflation, to be similar in all simulated cases. We arrive at the following polynomial form:
\begin{align}
    V_{\mathrm{Pert}}=\sum_{n=0}^4 a_n \phi^n,
\end{align}
where the perturbed coefficients are given in table~\ref{tab:Coefficients}. The last coefficient, $a_4$, is then recalculated to reset $\phi_{\mathrm{end}}=1$. As before, this is only a convenient way to choose perturbed inflationary potentials. The actual measure for perturbation of an inflationary potential is the adjusted $L_2$ norm such that in this case we have:
\begin{align}
    A_{\mathrm{Pert}}=\frac{\sqrt{\int \left(V_{\mathrm{Pert}}(\phi) - \left(1-\frac{1}{2\sqrt{2}+1}\phi^4\right)\right)^2 d\phi}}{\int \left(1-\frac{1}{2\sqrt{2}+1}\phi^4\right)d\phi}. 
\end{align}

After setting the functional form of the perturbed potential, we find the initial coordinate $\phi_0$ by requiring the number of efolds $N\gtrsim 65$. When assessing physical models, $\phi_{\mathrm{CMB}}$ is determined a-posteriori by finding the value of $\phi$ at, for instance, 60 efolds before the end of inflation. A very similar process is fully elaborated in \cite{Wolfson:2016vyx,Wolfson:2021utn}. However, in this case we set $\phi_{\mathrm{CMB}}\equiv 0$ for reasons explained previously. \\
{\renewcommand{\arraystretch}{2.5}
\begin{table}[]
    \centering
    \begin{tabular}{|| c | c ||}
        \hline
         $\displaystyle a_0$& $ \displaystyle 1$\\
         \hline
         $\displaystyle a_1$&$ \displaystyle \frac{2(a- 3b+3c-d)}{15.31 + a-1.5b+c-0.25 d}$\\ 
         \hline
         $\displaystyle a_2$& $\displaystyle \frac{-6(b - 2c +d)}{15.31 + a-1.5b+c-0.25 d} $\\
         \hline
         $\displaystyle a_3$& $\displaystyle \frac{8(c - d)}{15.31 + a-1.5b+c-0.25 d} $\\
         \hline
         $\displaystyle a_4$& $\displaystyle \frac{-4(1+d)}{15.31 + a-1.5b+c-0.25 d}\;\;^{(\star)}$\\
         \hline
    \end{tabular}\\
    \caption{Coefficients for a perturbed small field potential of the form $1-\phi^4$. $\{a,b,c,d\}$ are the fractional perturbations for the coefficients , where each is randomly chosen from a uniform distribution $U[-0.5,0.5]$. ($\star$) Finally, the coefficient $a_4$ is recalculated to yield $\phi_{\mathrm{end}}=1$.}
    \label{tab:Coefficients}
\end{table}}

In Figure~\ref{fig:SMF_fullAssymetry} we see that as little as a $0.1\%$ perturbation can yield over $1\%$ asymmetry in response. Moreover, even if the response is linear, fitting the data we find that above $\sim 0.5\%$ perturbation should consistently yield over $1\%$ asymmetry in response. The fluctuations in binned response is due to the relatively small number of cases simulated. We assume that with enough such simulations, the fluctuations eventually subside. This implies that, in the case of perturbed small field models, the predictive value of the analytic term is severely constricted.

\subsubsection{An Artificial Neural Network Approach to Small Field Models}
While the computational time required to calculate and retrieve the observables for one model is limited, it is nonetheless long enough to preclude incorporation into an MCMC engine and efficiently retrieving the posterior likelihood for a class of models. \\
In previous works we studied small-field models of a sixth degree polynomial given by:
\begin{gather}
    V(\phi)=V_0\left(1-\sqrt{\frac{r}{8}}\phi+\sum_{i=2}^{6}a_i\phi^i\right),\label{eq:6models}
\end{gather}
where the constraints of a fixed tensor-to-scalar ratio $r$, inflaton field excursion $\Delta\phi=1$ and efolding number $N=60$, leave three free parameters that fully define the model. The process of assigning likelihoods to models, is well documented in \cite{Wolfson:2016vyx,Wolfson:2018lel}. However, the process was long and required assumptions like a negligible paired covariance between cosmological observables. This in addition to resorting to trial and error made finding candidates extremely time consuming. In the future using a normalizing flow pipeline approach such as in \cite{Makinen:2021nly} may simplify the process.\\
\begin{figure}[t]
    \centering
    \includegraphics[width=0.85\textwidth]{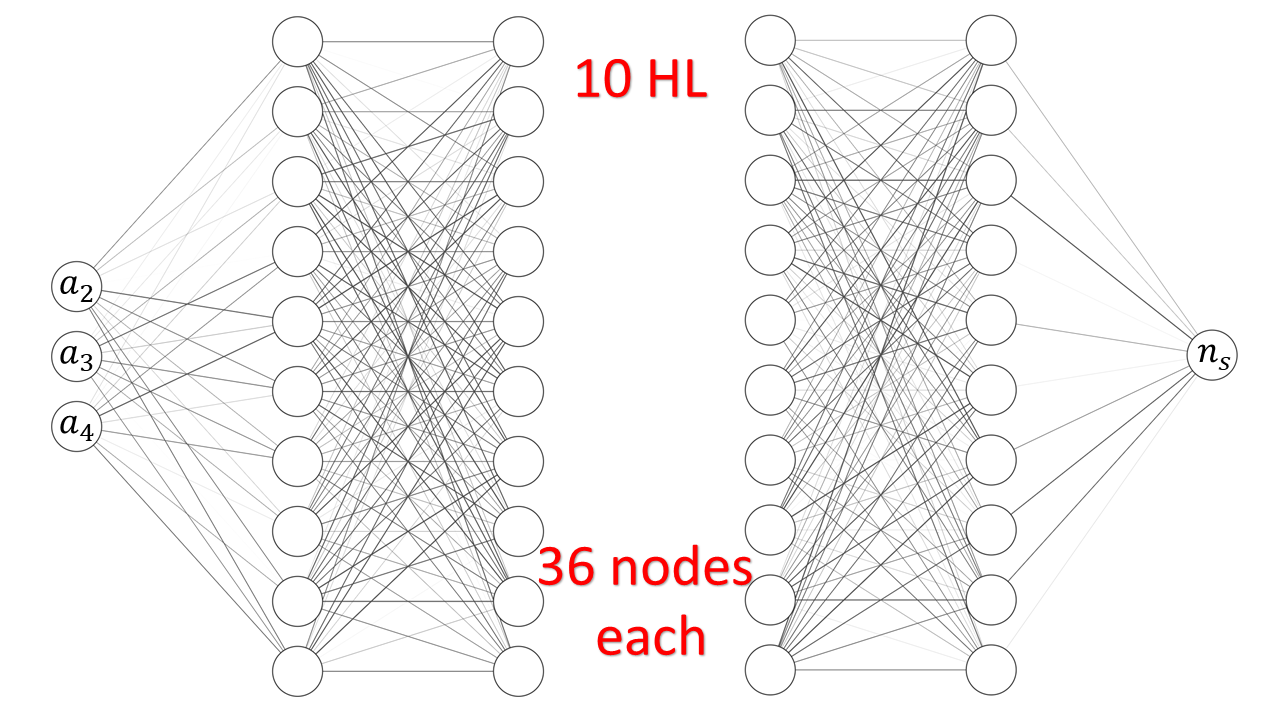}
    \caption{We employ increasingly intricate artificial neural networks to recover a stable connection between potential coefficients and the scalar index they yield. Ultimately the minimal architecture which yields sufficiently accurate predictions is comprised of 10 hidden layers with 36 nodes in each layer.}
    \label{fig:ANN}
\end{figure}\\
In this section, we aim to retrieve a correlation term between the potential shape and the scalar index. Specifically, we aim to find a function $\mathcal{F}$, such that for a potential of the form in \eqref{eq:6models}, $\mathcal{F}\left(a_1,a_2,a_3,a_4\right)=n_s$, consistently, up to $\sim0.1\%$ discrepancy. Finding such an expression, if simple enough, may re-enable the heuristic approach of small-field model-building. Additionally it will allow for sampling over potential coefficients, instead of sampling from the PPS observables, in a Bayesian analysis. Thus, the idea is, if we can find such a function for the types of models we look at, there is a possibility of finding a generalized correlation function. Conversely, failing to find such a function in these models indicates that in the functional space we probed, there is no good candidate. Thus, no good candidate for a {\textit{simple}} generalization exists, within this functional space. \\

In the models in \eqref{eq:6models}, asymmetry in $n_s$ between analytical approximations and numerical results were found at a level upwards of $1\%$.
Therefore we try to retrieve an analytical approximation $\mathcal{F}\left(a_1,a_2,a_3,a_4\right)=n_s$ by two methods: (1) Using multinomial fitting to recover a functional relation between coefficients and observables; And (2) using Artificial Neural Networks (ANNs) to look for patterns that correlate the models to their $n_s$ results swiftly and accurately.\\
Using the multinomial approach, we try to fit $\sim170,000$ simulated potentials with multinomials of up to $20^{\mathrm{th}}$ degree. We define a multinomial as an analytical multivariate function that can be written as:
\begin{gather}
    \mathcal{F}(a_2,a_3,a_4)=\sum_{n,l,m=0}^{D}b_{nlm}\left(a_2\right)^{n}\left(a_3\right)^{l}\left(a_4\right)^{m} \hspace{20pt}\big| n+l+m\leq D, \label{eq:multivariate}
\end{gather}
in which the degree of the multinomial $D$ is the highest possible power in $f$, and note that $a_1$ is constant in these models, thus it does not appear in the term above. This fitting is found to always yield a mean asymmetry upwards of $1\%$, and a maximum asymmetry upwards of $\sim 65\%$. In the ANN option, we use the TensorFlow framework \cite{abadi2016tensorflow}. We progressively use deeper and wider ANNs such that the input layer receives an ordered triplet of potential coefficients, $(a_2,a_3,a_4)$. The output layer yields the predicted $n_s$ value, as shown in Figure~\ref{fig:ANN}. We iterate over the number of hidden layers, and the number of nodes in each layer. The process is greedy such that the first (minimal) configuration of layers and nodes that conform to an accuracy threshold of $0.1\%$ maximum asymmetry, after training, is selected. Failing to do so, the output network will be the network with maximal mean accuracy in $n_s$ among the networks iterated over. Since the activation functions are all ReLU (Rectified Linear Unit), this netwrok models a generalized regressor to a multinomial function of degree no higher than the number of hidden layers $h$. Due to the greediness property, the selected multinomial model is of degree $h$ exactly.\\

We run this algorithm over the database of $\sim 170,000$ coefficients-observables samples, with $n_s$ ranging from as red as $n_s=0.3$ to as blue as $n_s=2.5$. The mean error of the ANN for predictions over the validation set is $\sim 1.5\%$, with no overfitting observed. The maximal error in $n_s$ over the validation set is $\sim 27\%$. We thus conclude that for a large phase space of observables, no reasonable analytical expression sufficiently approximates the resulting scalar index. Where by reasonable we mean a simple expression that can facilitate model building. This points to the map between coefficient phase-space to $n_s$ phase-space being extremely non-linear.\\

\begin{figure}[t]
    \centering
    \includegraphics[width=0.85\textwidth]{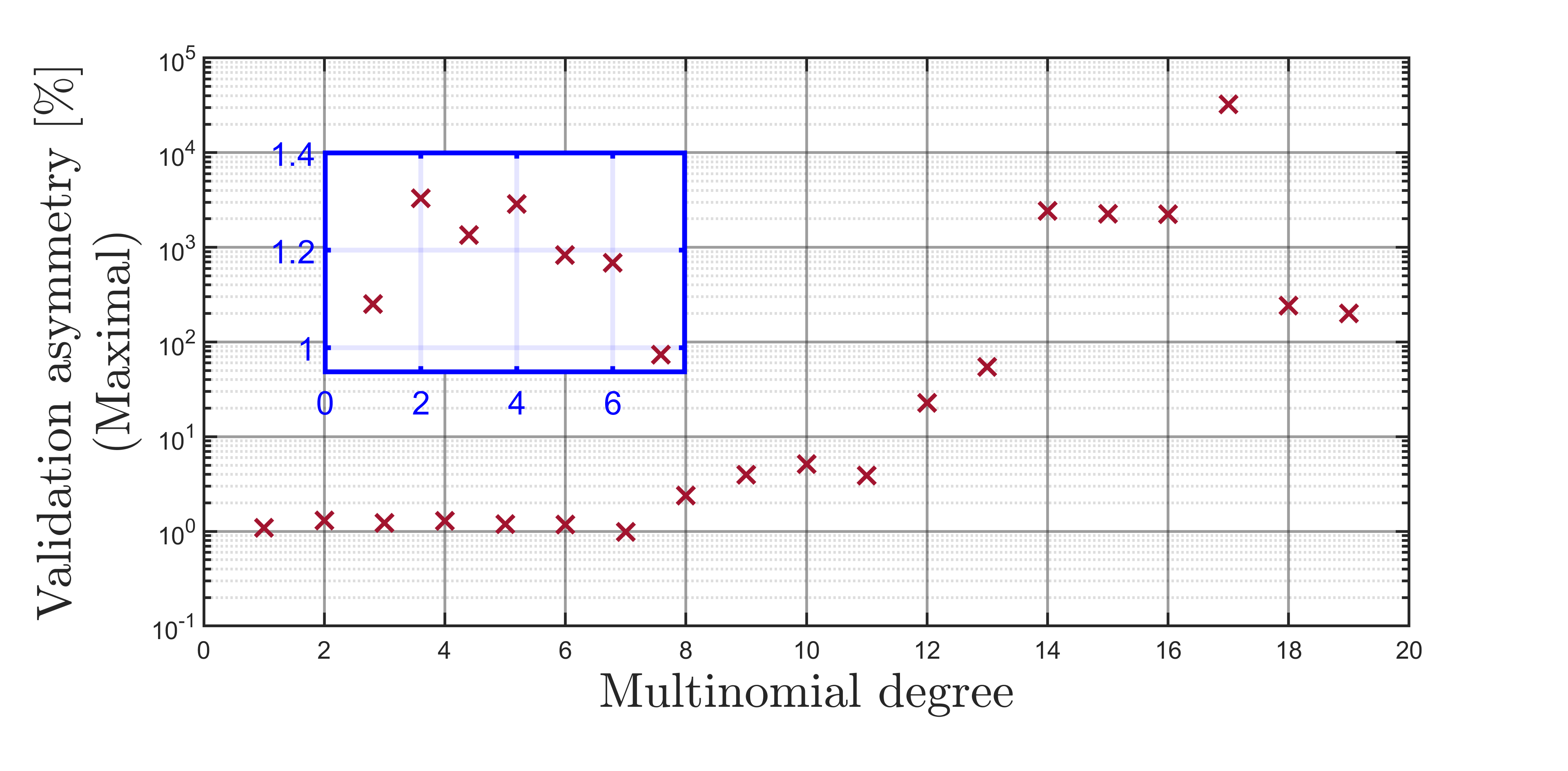}
    \caption{Using a subset of potentials that yield a calculated spectral index within the interval $n_s\in\left(0.955,0.985\right)$. The $7^{\mathrm{th}}$ degree multinomial seems to be the best fitting one. However, while the mean error is sufficiently small, the maximal error is still of the order of $1\%$. We can see that already at the $8^{\mathrm{th}}$ degree, overfitting appears.
    \label{fig:Multifit_approach}}
\end{figure}

Assuming a smooth mapping between phase-spaces, decreasing the target phase-space should simplify the functional mapping. We do this by including only models which yield a numerically calculated $n_s\in(0.955,0.985)$. This yields a subset of $\sim 2000$ cases.\\
We iterate over the multinomial degree, and record the mean and maximum asymmetries over a validation set. This method yields fitting terms of mean asymmetry of $\sim0.1\pm 0.07\%$ for a $7^{\mathrm{th}}$ degree multivariate function (as in \eqref{eq:multivariate}). This entails $120$ coefficients to fit.
However, the maximal asymmetry over the validation set, as shown in Figure~\ref{fig:Multifit_approach}, is always close or above $1\%$. This suggests a simple analytical term corresponding potentials of the small-field variety to PPS observables, with the desired less than $1\%$ maximal error, does not exist.\\

Applying the ANN approach to the $\sim 2000$ potential-result pairs data subset yields better results. For an ANN architecture that includes ten fully connected hidden layers of 36 nodes each, we recover a mean asymmetry of $\sim 0.01 \%$ and a maximum asymmetry of $\sim 0.5\%$, with no apparent overfitting. However, working with a constrained data set implies two pitfalls:
\begin{itemize}
    \item[1.] To get this data subset, we resorted to calculating and filtering a large sample of PPS yielded by inflationary potentials. This defeats the purpose of having such an analytic expression.
    \item[2.] The results of this ANN are naturally model dependant. 
\end{itemize}
\section{Discussion}
In this paper we studied the stability of the Mukhanov-Sasaki equation solutions with respect to perturbations over the inflationary potential. We first studied the PPS response to features of the inflationary potentials. Our intuition was that as the feature becomes sharper and larger in amplitude, the response becomes more pronounced. We looked at two types of features: 1) a step-like feature implemented by a hyperbolic tangent function, as it generates a delta-like feature in the pump-field; and 2) a Gaussian feature as a delta function generator over the baseline potential.\\

We found that indeed the analytic terms for the spectral index lose their predictive value in the limit of sharper features. Specifically we find that for a step-like feature spanning $\Delta\phi\sim 0.4-0.8\; $ the error in the analytic terms for the PPS are upwards of $\sim 100\%$. The same analysis for a Gaussian feature yielded an error upwards of $\sim 75\%$ with a feature of width  $\Delta\phi\sim 0.1-0.2\; $. While this type of analysis is of limited physical importance, it is valuable in promoting the understanding of the problem.\\

Following these findings we probed the validity limits of the first and second order analytic expressions, as a function of the perturbation measure. We set a consistent measure for perturbation by using a variant of the L2 norm (see eq.~\ref{eq:Pert_amp}). We found that for large field models, the predictions made by the Lyth-Riotto second order approximation are better than $1\%$ accuracy, up to a $\sim 3\%$ perturbation. But, for small field models a $0.1\sim 0.5\%$ perturbation is the limit of the $1\%$ accuracy mark. However, it does not mean perturbations of models beyond those limits are necessarily invalid. It simply implies that when working with highly perturbed potentials it is essential to check our results numerically.\\

Additionally, We have shown that a {\textit reasonable} general expression that accurately connects observables to small field potential coefficients, most likely does not exist. By {\textit reasonable} we mean in the sense that it doesn't promote a simple heuristic for model building. Even for a subset of potentials that yield the most likely spectral index of $n_s\sim0.965$ up to $\sim 2\%$, an expression that yields errors of order $\lesssim 1\%$ is found only at a seventh degree multinomial. \\

While applying an artificial neural network approach might seem promising, it includes the lengthy process of calculating the power spectra for a large sample of inflationary potentials. So it remains mostly unfeasible, until such time as setting up an unsupervised and efficient pipeline is possible.\\

It might be worth mentioning, that the analytical expressions for the index running $\alpha$, be it multinomial or ANN-derived, yield errors of the order of $10\sim100\%$, where the percentage is taken relative to the value of $n_s\sim 1$. However, failure at predicting $n_s$ has far reaching consequences, whereas the index running $\alpha$ is currently constrained to a lesser degree.\\

Thus, we supply a general `rule of thumb' for the validity of analytical terms correlating potentials to PPS observables in table~\ref{tab:Limits_Validity}. The process is the following:
\begin{itemize}
    \item[1.] Define your model and ascertain whether it is a LFM or SFM.
    \item[2.] Find the closest canonical potential by any curve fitting method. 
    For LFM's the canonical function would be a power-law inflation potential, and for SFM's the closest monomial potential.
    \item[3.] Use the measure in eq.~\ref{eq:Pert_amp}, to find the perturbation amplitude: 
    \begin{align}
    A_{\mathrm{Pert}}=\frac{\sqrt{\int\left(V_{\mathrm{Pert}}-V_{\mathrm{ref}}\right)^2 d\phi}}{\int V_{\mathrm{ref}}d\phi}, 
\end{align}
    where $V_{\mathrm{ref}}$ is the reference potential and $V_{\mathrm{Pert}}$ is the perturbed potential we intend to study.
    \item[4.] Refer to table~\ref{tab:Limits_Validity} to find the possible validity limits.
\end{itemize}
\begin{table}[!h]
    \centering
    \begin{tabular}{||c|c||}
        \hline
        Field excursion in the CMB window & Validity limit  \\
        \hline
        \hline
        $\Delta\phi >1$ & $\sim3\%$ \\
        $\Delta\phi<1$  & $0.1\sim 0.5\%$\\
        \hline
        \hline
    \end{tabular}
    \caption{Limits of validity for small field and large field models, in terms of perturbation percentage. This table only points at the limits for which we can be {\bf sure} the analytical expressions are valid to below $1\%$ error.}
    \label{tab:Limits_Validity}
\end{table}
We also demonstrate, that there is no other option except the numerical one in trying to correlate small field potentials of physical interest to the PPS they yield to better than $1\%$ accuracy. This is true regardless of the method used as the newer lattice calculations are obviously numerical in nature, and the Green's function approach relies on numerical integration. However, when studying a class of models rather than a single candidate, it is possible to employ an analytical approximation with a small enough mean error to recover the likelihood on model parameters by means of Bayesian analysis. Be that as it may, one may expect a posterior distribution markedly different from a simple Gaussian. This is due to the highly non-linear connection between observables and model parameters. In that sense, the problem of finding valid small field candidates of a certain class is made an inverse problem once again.\\

We believe this requires the model-building community to update their model-building toolkit. Since, if we want to make meaningful and testable predictions, the current analytical tools are insufficient. With the prospects of new CMB experiments, employing finer sensors this issue becomes all the more pressing.

\end{document}